\documentclass[a4paper,amsmath,amssymb,pre,twocolumn,superscriptaddress]{revtex4-1}

\usepackage{hyperref}
\usepackage{eepic}
\usepackage{graphicx}
\usepackage{overpic}
\usepackage{subfigure}
\usepackage{latexsym}
\usepackage{bm}% bold math
\usepackage{amsmath}
\usepackage{amssymb}
\usepackage{amsthm}
\usepackage{microtype}
\usepackage[retainorgcmds]{IEEEtrantools}

\def\kcore{$k$-core}
\def\ksubtree{$(k-1)$-ary subtree}
\def\kisubtree{$(k_i-1)$-ary subtree}
\def\hkcore{heterogeneous $k$-core}

\def\S{\mathcal{S}}
\def\C{\mathcal{C}}

\def\er{Erd\H{o}s-R\'enyi}
\def\etal{\textit{et al.}}
\def\ie{i.~e.}

\def\k{\mathbf{k}}

\begin{document}

\title{Critical phenomena in heterogeneous $k$-core percolation}
\author{Davide Cellai}
\affiliation{MACSI, Department of Mathematics and Statistics, University of Limerick, Ireland}
\author{Aonghus Lawlor}
%\affiliation{CBNI, University College Dublin, Belfield, Dublin 4, Ireland}
\affiliation{National Centre for Geocomputation, National University of Ireland, Maynooth, Co. Kildare, Ireland}
\author{Kenneth A.~Dawson}
\affiliation{CBNI, University College Dublin, Belfield, Dublin 4, Ireland}
\author{James P.~Gleeson}
\affiliation{MACSI, Department of Mathematics and Statistics, University of Limerick, Ireland}

\begin{abstract}
$k$-core percolation is a percolation model which gives a notion of network functionality and has many applications in network science.
In analysing the resilience of a network under random damage, an extension of this model isintroduced, allowing different vertices to have their own degree of resilience.
This extension is named heterogeneous $k$-core percolation and it is characterized by several interesting critical phenomena.
Here we analytically investigate  binary mixtures in a wide class of configuration model networks and categorize the different critical phenomena which may occur.
We observe the presence of critical and tricritical points and give a general criterion for the occurrence of a tricritical point.
The calculated critical exponents show cases in which the model belongs to the same universality class of facilitated spin models studied in the context of the glass transition.
\end{abstract}

\date{\today}
\maketitle

\section{Introduction}
A significant issue in many networked infrastructures is the threat that local damage can represent to the whole system \cite{newman2003,sterbenz2010}.
Moreover, there is an important distinction in the way a network collapses: it can happen in a smooth, predictable way, or with extreme abruptness.
The capability of designing networks where a collapse can be predicted in advance is expected to be increasingly important \cite{vespignani2010}.

In recent years, there has been a wide renewal of interest in percolation models, regarding the occurrence of a first order transition, instead of the classical continuous percolation transition \cite{achlioptas2009}.
However, it has been also shown \cite{bizhani2012} that discontinuous transitions are not so unusual in percolation models and can be dated to severals years ago \cite{ohtsuki1987,janssen2004}.
Moreover, several examples have been recently proposed in the context of explosive percolation \cite{araujo2010,cho2010,chen2011,manna2011}, interdependent networks \cite{parshani2010} and hierarchical lattices \cite{boettcher2012}.

%~ Perhaps the simplest notion of robustness of a network is the existence of a percolating cluster \cite{cohen2000}.
\kcore~percolation is an extension of percolation providing in a simple model a wide range of critical phenomena: a giant \kcore~cluster may collapse either continuously or discontinuously as a function of random damage \cite{Schonmann:1990p1812,branco1993,dorogovtsev2006}.
Moreover, the analysis of the \kcore~architecture of a network has developed into applications in different areas of science including protein interaction networks \cite{wuchty2005}, jamming \cite{schwarz2006}, neural networks \cite{Chatterjee2007}, granular gases \cite{alvarez2007}, evolution \cite{klimek2009}, social sciences \cite{kitsak2010} and the metal-insulator transition \cite{cao2010}.
Finally, a set of spin model approaches to the glass transition shares close similarities with \kcore~percolation \cite{sellitto2010}.

The analytical formalism of \kcore~percolation has been recently extended to include a local notion of robustness: some nodes can be more resilient than others and require a smaller number of neighbors to remain active.
This extension has been named \hkcore~(HKC) percolation  and has been studied in locally tree-like networks \cite{baxter2011}, finding a number of interesting critical phenomena including a tricritical point (TCP) \cite{cellai2011}.
%Quite interestingly, it has been shown that \kcore~percolation transitions can also be characterized by a discontinuous behavior \cite{branco1993,Schonmann:1990p1812}.
%

In this paper,  we extend the analysis of \cite{cellai2011} to present an exhaustive description of binary mixtures in \hkcore~percolation and show that \hkcore~percolation models are characterized by a wealth of critical behaviors, involving both first and second order transitions.
In Section 2 we define heterogeneous \kcore~percolation and explain the mathematical formalism in locally tree-like networks.
In Section 3 we focus on some illustrative examples of binary mixtures of vertex types, examining the different critical phenomena and calculating relevant critical exponents.
In Section 4 we give a general argument of the occurrence of a tricritical point in a binary mixtures of vertex types in heterogeneous \kcore~percolation.
Finally, Section 5 states our conclusions.

\section{The model}

% \subsection{$k$-core percolation}
Given a simple graph, a \kcore~is defined as the largest subgraph where every vertex has at least $k$ neighbors in the subgraph itself.
A common way of determining the \kcore~consists in removing recursively all the vertices (and adjacent edges) with less than $k$ neighbors.
We consider a framework where the nodes (and adjacent edges) of a given network are randomly removed with probability $1-p$ and we ask how the fraction of vertices in the \kcore~varies as $p$ is decreased.
An analytical formalism has been developed to study this problem on the configuration model \cite{dorogovtsev2006}.
The configuration model is defined as the maximally random network with a given degree distribution $P(k)$.
It has the important property that the number of loops vanishes as the size
$N\to\infty$, which guarantees that if a {\kcore} exists, it must be
infinite, at least if $k\geqslant2$ \cite{chalupa1979,dorogovtsev2006}.
This formalism is based on the definition of {\ksubtree}.
% \subsection{Heterogeneous $k$-core percolation}
Given the end of an edge, a
{\ksubtree} is defined as the tree where, as we traverse it, each vertex has at
least $k-1$ outgoing edges, apart from the one we came in.
In  configuration model networks, then, the \kcore~coincides with the \ksubtree.
The formalism by Dorogovtsev \etal~solves the problem in any locally tree-like
graphs and determines the order of the transition with which the giant
\kcore~collapses \cite{dorogovtsev2006}.

The {\hkcore} is an extension of the \kcore~where the minimum threshold $k_i$
is a local parameter, \ie~it depends on the vertex $i$.
The {\kisubtree}, then, is the tree in
which, as we traverse it, each encountered vertex has at least $k_i-1$
child edges.
We define $Z$ as the probability that a randomly chosen
vertex is the root of a {\kisubtree} and $M$ as the probability that a randomly
chosen vertex is in the HKC.
We can then write $M_{ab}$ for a mixture of two types of vertices $k_{a}, k_{b}$
where $k_{a}$ and $k_{b}$ vertices occur with probability $r$ and $1-r$,
respectively:
\begin{IEEEeqnarray}{rCl}
  M_{ab}(p)  &=&  \bar{M}_a(p) + \bar{M}_b(p)
  \label{eq:Mab}
\end{IEEEeqnarray}
\begin{equation}
  \bar{M}_a(p) = p r\sum_{q=k_a}^{\infty} P(q) \sum_{l=k_a}^{q} \Phi_{l,q}(Z,Z)
  \label{eq:Ma_bar}
\end{equation}
\begin{equation}
  \bar{M}_b(p) = p (1-r)\sum_{q=k_b}^{\infty} P(q) \sum_{l=k_b}^{q}\Phi_{l,q}(Z,Z)
  \label{eq:Mb_bar}
\end{equation}
where $\bar{M}_{a(b)}(p)$ is the fraction of nodes of type $a(b)$ in
the \hkcore, respectively, $P(q)$ is the degree distribution of the network and we
have used the convenient auxiliary function:
\begin{equation}
  \Phi_{l,q}(X,Z) = {q \choose l} (1-Z)^{q-l} \sum_{m=1}^l {l \choose m} X^m
(Z-X)^{l-m} \nonumber.
\end{equation}
It is shown in \cite{baxter2011} that $Z$ satisfies the self-consistent
equation:
\begin{IEEEeqnarray}{rCl}
	Z &=& p r \sum_{q=k_a}^{\infty} \case{qP(q)}{\langle q \rangle}
\sum_{l=k_a-1}^{q-1} \Phi_{l,q-1}(Z,Z) +\nonumber\\
	&& + p(1-r)\sum_{q=k_b}^{\infty} \case{qP(q)}{\langle q \rangle}
\sum_{l=k_b-1}^{q-1} \Phi_{l,q-1}(Z,Z).
	\label{eq:Z-equation}
\end{IEEEeqnarray}

In general, it is not granted that the \hkcore~is uniquely made up of a giant
cluster.
As it is possible that some finite clusters of the \hkcore~are present, Baxter
\etal~develop the equation for $X$, the probability that an arbitrarily
chosen edge leads to a vertex which is the root of an \emph{infinite}
\kisubtree.
In the case of a binary mixture, the equation reads
\begin{IEEEeqnarray}{rCl}
	X &=& pr \sum_{q=k_a}^{\infty} \case{qP(q)}{\langle q \rangle}
\sum_{l=k_a-1}^{q-1} \Phi_{l,q-1}(X,Z) +\nonumber\\
	&& +p(1-r)\sum_{q=k_b}^{\infty} \case{qP(q)}{\langle q \rangle}
\sum_{l=k_b-1}^{q-1} \Phi_{l,q-1}(X,Z),
	\label{eq:X-equation}
\end{IEEEeqnarray}

Therefore, the probability $\S_{ab}$ that a randomly chosen vertex belongs to
the \emph{giant} HKC is
\begin{equation}
  \S_{ab} = \sum_{l\ge k_a}\S_a(l) + \sum_{l\ge k_b}\S_b(l)
  \label{eq:S-equation}
\end{equation}
with
\begin{equation}
  \S_a(l) = p r \sum_{q\ge l} P(q) \Phi_{l,q}(X,Z),
\end{equation}
\begin{equation}
  \S_b(l) = p (1-r) \sum_{q\ge l} P(q) \Phi_{l,q}(X,Z).
\end{equation}
$\S_{a(b)}(l)$, then, is the probability that an $a(b)-$vertex in the giant HKC has exactly $l\ge k_{a(b)}$ neighbors in the giant HKC, respectively.
Those expressions are analogous to (\ref{eq:Ma_bar}) and (\ref{eq:Mb_bar}), as can be seen after using the identity $\sum_{q=k}^{\infty}\sum_{l=k}^{q}c_{q,l} = \sum_{l=k}^{\infty}\sum_{q=l}^{\infty}c_{q,l}$, valid for any function $c_{q,l}$.

An important concept in {\kcore} percolation is the \emph{corona}.
The corona is defined as the subset of the HKC where every vertex $i$ has
exactly $k_i$ nearest neighbours in the HKC.
By definition, then, a corona cluster is characterized by the property that if
only one of its vertices is removed, the whole cluster collapses.
It can be shown that the corona clusters are finite everywhere except at the
phase transition, where the mean cluster size diverges
\cite{goltsev2006,baxter2011}.
Using equation (\ref{eq:S-equation}), then, the corona $\C_{ab}$ of a binary
mixture is given by the following formula
\begin{eqnarray}
	\C_{ab} & = & \S_a(k_a) + \S_b(k_b) = \nonumber\\
	& = & p r \sum_{q\ge k_a} P(q) \Phi_{k_a,q}(X,Z)  +\nonumber\\
	&&  + p (1-r) \sum_{q\ge k_b} P(q) \Phi_{k_b,q}(X,Z).
	\label{eq:corona}
\end{eqnarray}

The distinction between \hkcore~clusters and the giant \hkcore~has to be made
whenever the two do not coincide.
That is the case when we consider binary mixtures of the type $k_a=1$, $k_b>1$,
as for $k=1$ finite 1-core clusters are possible.
However, for mixtures such as $2\leq k_a < k_b$, no finite \hkcore s are
possible, therefore: $X=Z$ and $M_{ab} = \S_{ab}$.

As in the homogeneous case, also in heterogeneous \kcore~percolation a
\kcore~architecture of the network can be defined.
Therefore we define a \emph{heterogeneous sub-core} as a subset of the HKC where a higher threshold $k_i$ is imposed on some or all vertex types.
More specifically, given a HKC of type $k_a$, $k_b$, the strength $M_k$ of a
heterogeneous sub-core is given by
\begin{eqnarray}
%	M_k & = & \mathcal{M}^a_k + \mathcal{M}^b_k = \nonumber\\
	M_k 	& = & p r \sum_{l\ge h_a} \sum_{q\ge l} P(q) \Phi_{l,q}(Z,Z) + \nonumber\\
		& & + p (1-r) \sum_{l\ge h_b} \sum_{q\ge l}  P(q) \Phi_{l,q}(Z,Z),
	\label{eq:M-subcore-definition}
\end{eqnarray}
where $h_a=max(k,k_a)$ and $h_b=max(k,k_b)$.
An analogous expression holds for $\S_k$:
\begin{eqnarray}
	\S_k & = & p r \sum_{l\ge h_a} \sum_{q\ge l} P(q) \Phi_{l,q}(X,Z) + \nonumber\\
		& & + p (1-r) \sum_{l\ge h_b} \sum_{q\ge l}  P(q) \Phi_{l,q}(X,Z).
	\label{eq:S-subcore-definition}
\end{eqnarray}
Therefore, when considering for example a binary HKC with $k_a <k_b$,
sub-cores with $k\le k_a$ coincide with the HKC, whereas sub-cores
with $k > k_a$ are proper subsets of the HKC.

\section{Illustrative examples}

We consider now a few examples of binary mixtures on \er~graphs with the most
interesting critical phenomena.
Plugging in the Poissonian degree distribution $P(q) = z_1^q \exp(-z_1)/q!$ (where
$z_1$ is the mean degree) into equations (\ref{eq:Z-equation}) and
(\ref{eq:X-equation}), the sums can be calculated analytically and several
critical phenomena can be found.
First, we review two cases already studied in the literature: the case
$\k=(k_a,k_b)=(1,3)$, examined in \cite{baxter2011} on the Bethe lattice, and the case
$\k=(2,3)$ \cite{cellai2011}.
Then, we focus on the case  $\k=(2,4)$, and show that its phase diagram
is qualitatively identical to the ones of the type $k_a=1$, $k_b\ge 3$, and
represents a  quite general behavior, with applications to the theory of glass transitions.

\subsection{The case $k_a=1$, $k_b=3$}

%\subsubsection{Solution of the equations}
In this case we have to take into consideration that, due to the presence of
vertices of type 1, there are finite $k$-cores.
The two equations (\ref{eq:Z-equation}) and (\ref{eq:X-equation}) can be
re-written as
\begin{eqnarray}
	p f_{13}(Z) &=& 1\\
	p h_{13}(X,Z) &=& 1,
\end{eqnarray}
where
\begin{equation}
	f_{13}(Z) = \frac{1 - (1-r)e^{-z_1 Z}(1 + z_1Z)}{Z},
\end{equation}
\begin{equation}
	h_{13}(X,Z) = \frac{1-e^{-z_1 X}}{X} -(1-r)z_1 e^{-z_1 Z}.
\end{equation}
It can be shown that the locus defined by $f_{13}'(Z)=0$ corresponds to a line of first order transitions which ends into a critical point defined by $f_{13}'(Z)=f_{13}''(Z)=0$  \footnote{We comment further about the details of the solution in the case $\k=(2,4)$, where the phase diagram is qualitatively the same.}.
A simple calculation yields:
\begin{equation}
	r_c = 1 - \frac{1}{3}e
\end{equation}
\begin{equation}
	p_c = \frac{3}{z_1}
\end{equation}
\begin{equation}
	Z_c = \frac{1}{z_1}
\end{equation}
$X_c$ has no analytic expression, being the non-trivial solution of the equation
$1-\exp(-z_1X_c) = 2/3 z_1X_c$, which is $z_1X_c \simeq 0.874$.

%\subsubsection{Strengths of percolating clusters}
Using (\ref{eq:S-equation}), we can calculate the strength of the giant HKC as
\begin{eqnarray}
	\S_{13} &=& p\left[ 1 - e^{-z_1X}\right] + \nonumber\\
	 && -p(1-r)z_1X e^{-z_1 Z}\left(1+z_1Z-\frac{1}{2}z_1X\right),
\end{eqnarray}
and we can also derive the expression of the 2-sub-core (from eq.~(\ref{eq:S-subcore-definition})):
\begin{eqnarray}
	\S_{2} &=& p r (1 - e^{-z_1X} -z_1X e^{-z_1 Z}) + \nonumber\\
	&& +  p(1-r) \left[ 1 - e^{-z_1X} +\right. \nonumber\\
	&& \left. - z_1X e^{-z_1 Z}\left(1+z_1Z-\frac{1}{2}z_1X\right) \right].
\end{eqnarray}
Using (\ref{eq:corona}), the strength of the HKC corona is given by the following expression:
\begin{eqnarray}
	\C_{13} &=& p e^{-z_1 Z} \left[ rz_1X + \right.\nonumber\\
	&& \left. +\frac{1}{6}(1-r)z_1^3\left(
3Z^2X-3ZX^2+X^3 \right)  \right].
\end{eqnarray}
In particular, the fraction of vertices of each type in the HKC are given by the
following expressions:
\begin{equation}
	\bar{\S}_{1} = p r (1 - e^{-z_1X})
\end{equation}
\begin{equation}
	\bar{\S}_{3} = p(1-r)\left[1 - e^{-z_1X} - z_1X e^{-z_1
Z}\left(1+z_1Z-\frac{1}{2}z_1X\right)  \right]
\end{equation}

The critical percolating strengths can be calculated with arbitrary precision. The approximate values are given in Table I. %\ref{table:13-critical}.
\begin{table}[htb]
\begin{tabular}{ccccc}
\hline \hline
 $r_c$ & $z_1\S^c_{13}$ & $z_1\C^c_{13}$ & $z_1\bar{\S}^c_{1}$ & $z_1\bar{\S}^c_{3}$ \\
\hline
 0.0939 & 0.3821 & 0.2700 & 0.1642 & 0.2179\\
\hline \hline
\end{tabular}
\label{table:13-critical}
\caption{Critical fractions of some relevant quantities in the case $\k=(1,3)$.}
\end{table}

%\subsubsection{Phase diagram}

\begin{figure}[htb]
	\begin{center}
%	\begin{overpic}[width=0.99\columnwidth,angle=0,grid=5]{figures/1-3-phase-diag-Mr}
%		\put(60,60){
%	    \includegraphics[width=0.28\columnwidth,angle=0]{figures/1-3-phase-diag-pr}}
%	%		\includegraphics[width=0.99\columnwidth,angle=0]{figures/1-3-phase-diag-Mr}
%	\end{overpic}
	 \includegraphics[width=\columnwidth]{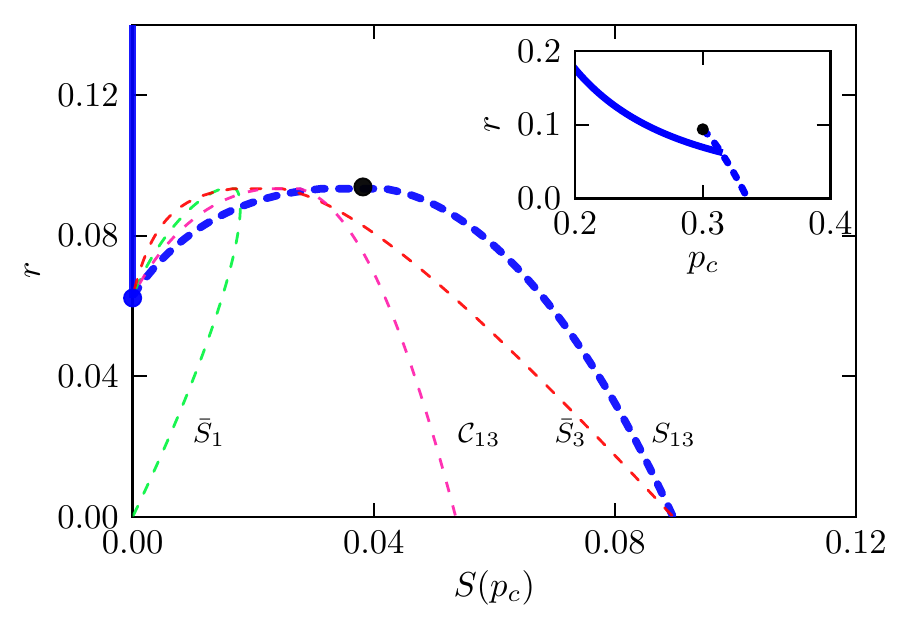}
	\caption{Phase diagram of the case $\k=(1,3)$ for the \er~graph ($z_1=10$). 
	The main panel illustrates the critical values of the giant HKC strength.
	The blue lines report the second order (solid) and the first order (dashed) transitions.
	The critical (black) and the critical end point (blue) are indicated by a dot.
	The fractions of nodes of type 1 ($\bar{\mathcal{S}}_1$) and 3 ($\bar{\mathcal{S}}_3$) in the giant HKC are also reported.
	The corona size $\mathcal{C}_{13}$ is represented by a magenta dashed line.
	The inset reports the most relevant detail of the phase diagram in the space ($r$,$p_c$).
	}
	\label{fig:1-3-ph-diag}
\end{center}
\end{figure}

Figure \ref{fig:1-3-ph-diag} displays the  phase diagram of the mixture $\k=(1,3)$ for a  \er~graph.
The two phase region occurs at a relatively low value of $r_c$, meaning that many 3-nodes are necessary to drive the system towards the first order (hybrid) transition.
The relative composition of the giant HKC, instead, presents a much higher fraction of 1-nodes, due to the high fragility of 3-nodes with respect to 1-nodes.
The phase diagram also shows that the corona strength is much smaller than $\S_{13}$ on the right side of the coexistence region, whereas is much closer to $\S_{13}$ on the left side, showing that the corona clusters dominate the HKC in the 1-rich phase.

 %~ \begin{figure}[htb]
%~ \includegraphics[width=0.99\columnwidth,angle=0]{figures/1-3-phase-diag-corona}
 %~ \caption{
 		%~ Phase diagram of the case (1,3) for the \er~graph ($z_1=10$).
 		%~ }
 %~ \label{fig:1-3-ph-diag-corona}
 %~ \end{figure}

\subsection{The case $k_a=2$, $k_b=3$}

As shown in our recent paper, this case has the property $X=Z$ and is characterized by a tricritical point \cite{cellai2011}.
We review the peculiarities of this case.
The quantity $Z$ at a given damage fraction $(1-p)$ can be calculated by solving the equation $pf_{23}(Z)=1$, where
\begin{equation}
	f_{23}(Z) = \frac{1-e^{-z_1Z}[1+(1-r)z_1Z]}{Z}.
\end{equation}
From (\ref{eq:Mab}), the strength of the \hkcore~is given by
\begin{equation}
	M_{23}(p) =
p\left\{1-e^{-z_1Z}\left[1+z_1Z+\frac{1}{2}(1-r)z_1^2Z^2\right]\right\}.
\end{equation}
It is easy to see that $f_{23}(Z)$ has a maximum at a finite $Z$ for $r<1/2$, which continuously moves to $Z=0$ at exactly $r=1/2$.
This implies that there is a line of first order transitions for $r<1/2$, and a second order percolating transition for $r>1/2$.
At $r=1/2$, the two lines match exactly at a tricritical point (TCP).
The critical exponent $\beta$ at the transition $M_{23}(p) - M_{23}(p_c) \sim (p-p_c)^{\beta}$, can be calculated analytically:
\begin{equation}
\beta = \left\{%
\begin{array}{lcrcl}
2 &&1/2 \leqslant &r& < 1\\
1 &&&r&=1/2\\
1/2  &&0 \leqslant &r&< 1/2
\end{array}
\right.
\end{equation}
The value of the exponent $\beta$ at  $r< 1/2$ agrees with the typical hybrid transition phenomenology \cite{dorogovtsev2006,branco1993}, whereas for $r> 1/2$ we recover the exponent of classical percolation without dangling ends \cite{Schonmann:1990p1812}.
The value of $\beta$ at the TCP  does not change when it is calculated along a line at fixed $p_c$.

Another exponent which can be calculated analytically is the one which governs the vanishing of the coexistence region as $r\to 1/2^-$.
We indicate it as $M^*(r) \sim \left(\frac{1}{2}-r\right)^{\beta_u}$, with $\beta_u=2$.
Finally, the rotation defining the critical fields is
\begin{equation}
\left(
\begin{array}{c}
\mu_{\perp} \\
\mu_{\parallel}
\end{array} \right)
=
\left(
\begin{array}{cc}
\cos\theta & \sin\theta\\
-\sin\theta & \cos\theta \\
\end{array} \right)
\left(
\begin{array}{c}
p-p_t \\
r-r_t
\end{array} \right)
\end{equation}
with $\tan\theta = 4/z_1$.
Close to the tricritical point, the critical line has
a behavior $\mu_{\parallel}\sim \mu_{\perp}^{1/2}$, with a crossover
exponent $\varphi_t=2$.
%Fig.~\ref{fig:2-3-crossover} shows that the numerically calculated exponent matches quite well the value calculated analytically.
%~ 
%\begin{figure}
%\includegraphics[width=0.95\columnwidth,angle=0]{figures/er-2-3-p-r-crossover}
%\caption{Behavior of the new coordinates close to the tricritical point.}
%\label{fig:2-3-crossover}
%\end{figure}

\begin{figure}[bht!]
	\begin{overpic}[width=\columnwidth]{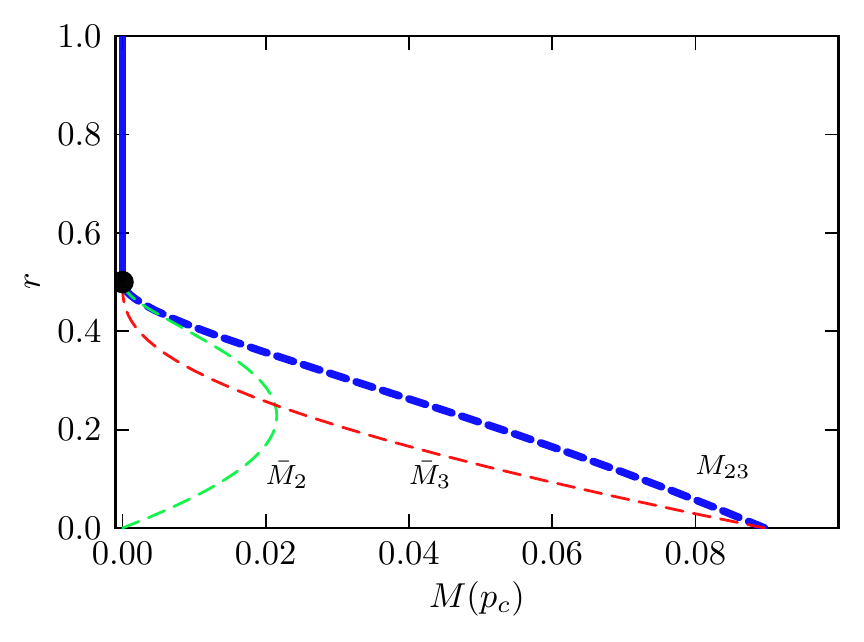}
		\put(59,27){\includegraphics[height=0.4\columnwidth,angle=0,keepaspectratio=true]{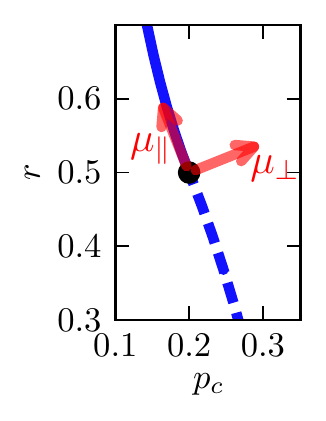}
		}
	\end{overpic}
	\caption{Phase diagram of the $\k=(2,3)$ mixture, showing the
          total mass of the percolating {\hkcore} cluster at different
          compositions $r$, for {\er} networks with $z_{1}=10$. The TCP
          at $r=1/2$ separates a line of first order transitions
          (dashed) from the second order line (solid). The masses of the fractions of nodes of types 2 (green) and 3 (red) in the giant HKC are also shown.
          The inset shows the phase diagram in the $(r,p)$ space.}
	\label{fig:ph_diag_2_3}
\end{figure}

Fig.~\ref{fig:ph_diag_2_3} shows the phase diagram of this case.
The system undergoes a first order (hybrid) phase transition for $r<1/2$, which smoothly shrinks at the TCP.
The fractions $\bar{M}_2$ and $\bar{M}_3$ of nodes of each type inside the HKC are also plotted.
It is interesting to note that $\bar{M}_2$ is substantially higher than $\bar{M}_3$ in the proximity of the TCP.
This behavior, which is the opposite of what occurs in the case $\k=(1,3)$, appears to be related to the higher value of $r$ at which the TCP occurs: the fraction of nodes of type 2 in the critical HKC steadily grows along the first order line as the coexistence region shrinks.
In the case $\k=(1,3)$, instead, the critical point occurs at quite a low $r$, due to the high stability of type 1 nodes, before the switch between the two mixture components.
It appears that the vanishing of the discontinuity is finely tuned by the vanishing fraction of type 3 nodes in the HKC.
At the same time, of course, the fraction of type 2 nodes becomes increasingly important, determining the change of order of the transition at the TCP.
%~ 

\subsection{The case $k_a=2$, $k_b=4$}

%\subsubsection{Introduction}

From equation (\ref{eq:Z-equation}), the case $k_a=2$, $k_b=4$ is solved by $pf_{24}(Z) = 1$, where
\begin{equation}
 f_{24}(Z) = \frac{ 1 - e^{-z_{1}Z} \left\{ 1+ (1-r) \left( z_{1} Z
+\frac{1}{2} z_{1}^2 Z^2\right) \right\} }{Z}.
\label{eq:2-4-fZ}
\end{equation}

%\begin{figure}[htb!]
%  \includegraphics[width=0.95\columnwidth]{figures/2-4-fZ}
%  \caption{$f(Z)$ in the case $k_a=2$, $k_b=4$ for some values of $r$
%($z_1=10$).}
%  \label{fig:2-4-fZ}
%\end{figure}%

\begin{figure}[htb]
\begin{minipage}[]{0.49\columnwidth}
	\subfigure[$r=0$]{%
		\includegraphics[width=\columnwidth,angle=0]{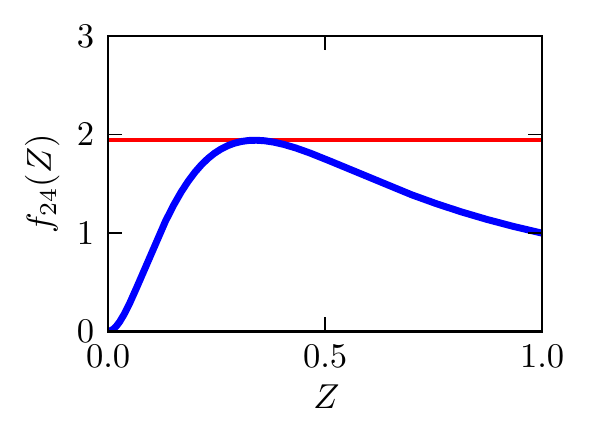}
	}
%	\label{fig:subfig:2-4-fZ-1}
\end{minipage}
\begin{minipage}[]{0.49\columnwidth}
	\subfigure[$r=0.15$]{%
		\includegraphics[width=\columnwidth,angle=0]{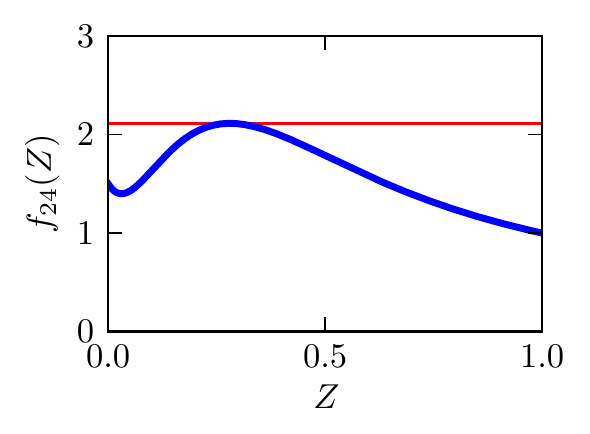}
	}
\end{minipage}
\begin{minipage}[]{0.49\columnwidth}
	\subfigure[$r=r_{cep}$]{%
		\includegraphics[width=\columnwidth,angle=0]{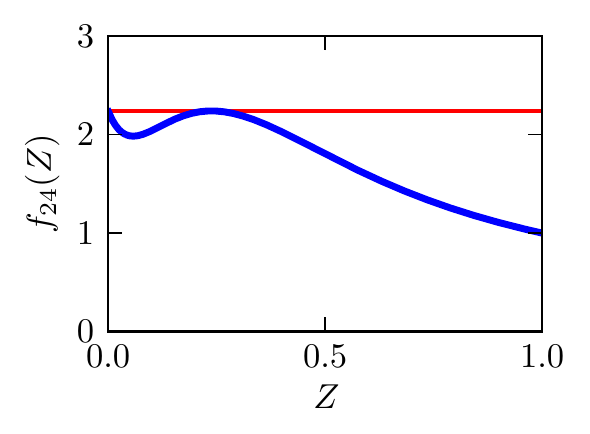}
	}
\end{minipage}
\begin{minipage}[]{0.49\columnwidth}
	\subfigure[$r=0.25$]{%
		\includegraphics[width=\columnwidth,angle=0]{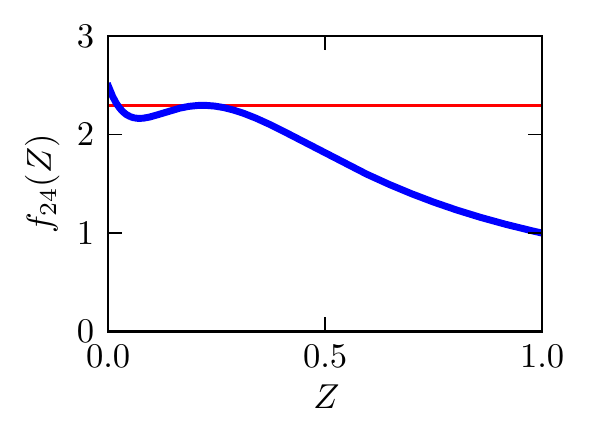}
	}
\end{minipage}
\begin{minipage}[]{0.49\columnwidth}
	\subfigure[$r=r_{c}$]{%
		\includegraphics[width=\columnwidth,angle=0]{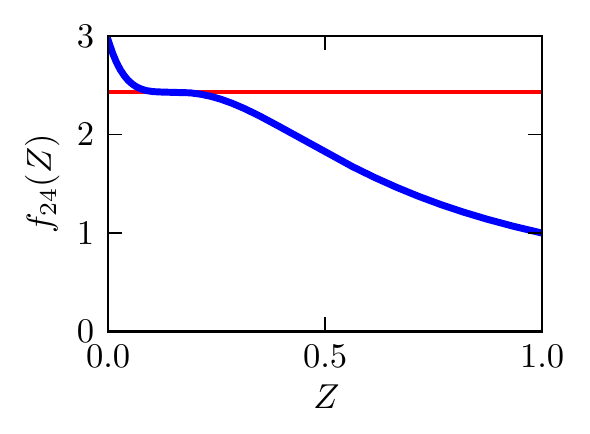}
	}
\end{minipage}
\begin{minipage}[]{0.49\columnwidth}
	\subfigure[$r=0.35$]{%
		\includegraphics[width=\columnwidth,angle=0]{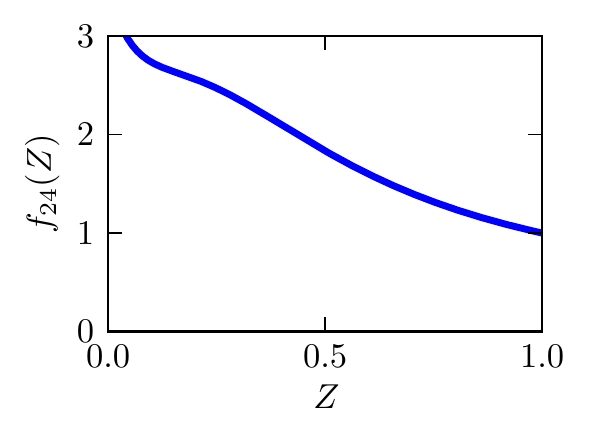}
	}
\end{minipage}
\caption{Different scenarios of the solutions of equation $pf_{24}(Z)=1$ in the case $k_a=2$, $k_b=4$ for different values of $r$ ($z_1=10$). The red horizontal line marks the value of $1/p_c$ where the first order transition occurs. In the panels (a) and (b) there is a first order transition corresponding to the global maximum of $f_{24}(Z)$; in (c) the maximum is at the same height as $f_{24}(Z=0)$ (critical end point); in (d) there is both a first and a second order transition; in (e) the first order transition disappears into a critical point; in (f) only the classical percolation transition at $Z=0$ remains.}
\label{fig:2-4-fZ}
\end{figure}

Differently from the case $k_a=2$, $k_b=3$, here for some values of $r$ the
function $f_{24}$ has a local maximum which does not approach the $Z=0$ point.
More specifically, it is easy to show that $f_{24}(Z)$ has
three types of behavior: monotonously decreasing, a local maximum at $Z>0$, and
a global maximum at $Z>0$ (Fig.~\ref{fig:2-4-fZ}).
A global maximum at $Z=0$ corresponds to a second order de-percolating transition, as the solution of equation $p f_{24}(Z)=1$ smoothly vanishes at some $p_c$.
A local maximum at $Z>0$, instead, determines a first order transition line
between two stable giant HKCs.

% \subsubsection{Masses}

Using equation (\ref{eq:Mab}), the strength of the HKC can be re-written as
\begin{eqnarray}
	M_{24}(p) &= & p \left\{1-e^{-z_{1}Z}\left[1+z_{1}Z
+\frac{1}{2}(1-r)z_{1}^2 Z^2  +\right.\right.\nonumber\\
	&& + \left. \left.\frac{1}{6} (1-r)z_{1}^3 Z^3\right]  \right \} .
\end{eqnarray}
From the definition (\ref{eq:M-subcore-definition}), the strength of the heterogeneous sub-core as a subset of the HKC is given by
\begin{equation}
  M_k(p) = \mathcal{M}^a_k + \mathcal{M}^b_k,
  \label{eq:2-4-subcore}
\end{equation}
where
\begin{equation}
 \mathcal{M}^a_k = pr \left[ 1 - e^{-z_1Z}\sum_{n=0}^{h_a-1}\frac{(z_1Z)^n}{n!}
\right]
\end{equation}
\begin{equation}
 \mathcal{M}^b_k = p(1-r) \left[ 1 -
e^{-z_1Z}\sum_{n=0}^{h_b-1}\frac{(z_1Z)^n}{n!}
\right]
\end{equation}
and $h_{a(b)}=max(k,k_{a(b)})$.
For $k\le 2$, the sub-core coincides with the HKC and for instance we have $M_2 =
M_{24}$.
For $k\ge 4$, the sub-core is a proper subset of the HKC, but the vertex
distinction is not longer relevant, as the threshold $k$ applies in the same
way to both vertex types.
Therefore, the sub-core coincides with the usual homogeneous \kcore.
The most interesting case is when $k=3$, as this threshold restricts the number
of acceptable type 4 vertices, but it has no effect on the threshold of type 2
vertices.
We indicate the corresponding strength as $M_{3}$, which is given by
the following formula from (\ref{eq:2-4-subcore}):
\begin{eqnarray}
	M_{3}(p) &=& p \left\{1-e^{-z_{1}Z}\left[1+z_{1}Z +\frac{1}{2} z_{1}^2
Z^2 \right.  \right. +\nonumber\\
	&& + \left.\left. \frac{1}{6} (1-r)z_{1}^3 Z^3 \right]  \right\} .
\end{eqnarray}
%Finally, from (\ref{eq:corona}) the corona of the HKC is
%\begin{equation}
%	C_{24}(p) = p e^{-z_{1}Z} \left\{\frac{1}{2} rz_{1}^2 Z^2  +\frac{1}{4!}
%(1-r)z_{1}^4 Z^4  \right \} .
%\end{equation}

%\subsubsection{Phase diagram}

\begin{figure}[htb]
\begin{center}
	\includegraphics[width=\columnwidth]{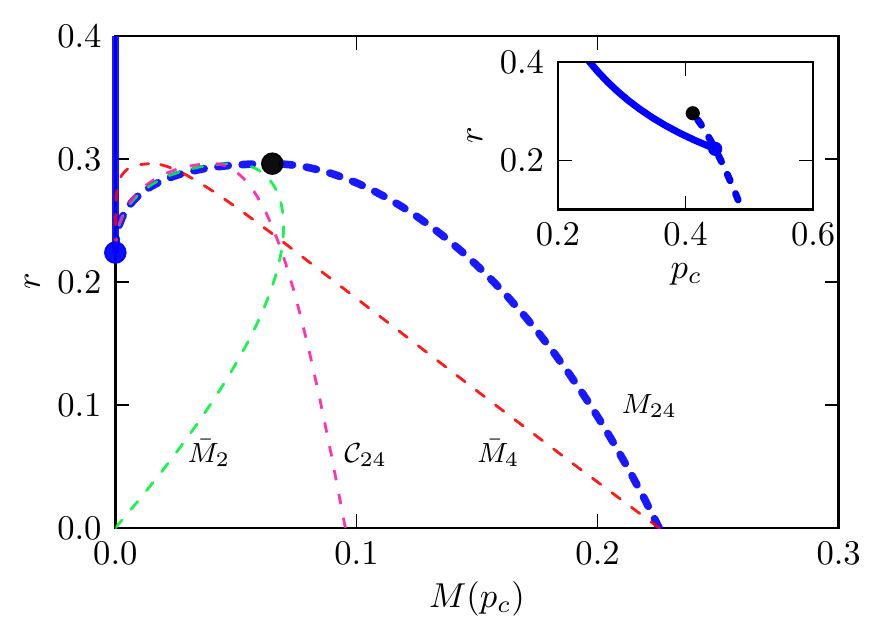}
%	\begin{overpic}[width=0.99\columnwidth,angle=0,grid=5]{figures/2-4-phase-diag-Mr}
%		\put(60,60){\includegraphics[width=0.3\columnwidth,angle=0]{figures/2-4-phase-diag-pr}}
%	\end{overpic}
	\caption{Phase diagram of the case $\k=(2,4)$ for the \er~graph ($z_1=10$) in the space ($r$,$M$), main panel, and ($r$,$p$), inset. 
	Symbols are as in Fig.~\ref{fig:1-3-ph-diag}}.
	\label{fig:er-2-4}
\end{center}
\end{figure}

From (\ref{eq:2-4-fZ}), we calculate the phase diagram at different
compositions $r$ (Fig.~\ref{fig:er-2-4}).
At high $r$, the phase diagram is characterized by a critical line of
de-percolating transitions. 
This line meets a first order line at a point which is usually called a
\emph{critical end point} in condensed matter physics.
The first order line corresponds to the type of hybrid transition observed in
\kcore~percolation for $k\ge 3$.
Differently from the case $k_a=2$, $k_b=3$, here the first order line
presents a \emph{critical point} at the end of a two-phase coexistence between a low
and a high density phase.

The position of the critical end point is determined by imposing $f_{24}'(Z)=0$ and $f_{24}(Z)=rz_1$, which yield the  solution $r_{cep} = 0.2239964831566\dots$.
Now we consider the strength $M_{24}^*(r)$ of the HKC along the low density border of the coexistence region in approaching the critical end point.
We define the critical exponent $\beta_{cep}$ of the critical end point from the manner in which such low density border vanishes: $M_{24}^*(r)\sim(r-r_{cep})^{\beta_{cep}}$.
It emerges that as $r\to r_{cep}^+$ we have:
\begin{equation}
	M_{24}^*(r) \sim (r-r_{cep})^2,
\end{equation}
which implies that $\beta_{cep}=2$.
The strength of the $3$-sub-core in the HKC can be calculated in a similar way
\begin{equation}
	M_{3}^*(r) \sim (r-r_{cep})^3.
\end{equation}
%
%\begin{equation}
%	M_{24}^*(r) = \frac{ [2-e^{-x^*}(2+x^*)]^2 }{ 2z_1 {r_{cep}}^2 } (r-r_{cep})^2 + o[ (r-r_{cep})^2],
%\end{equation}
%where
%\begin{equation}
%	x^* = \frac{(1-r_{cep})^2 + \sqrt{1-4r_{cep}-2r_{cep}^2+4r_{cep}^3+r_{cep}^4}}{2r_{cep}(1-r_{cep})} .
%\end{equation}
%This implies that $\beta_{cep}=2$.
%
%
%The strength of the $3$-sub-core in the HKC can be calculated in a similar way
%\begin{eqnarray}
%	 M_{3}^*(r) &=& \frac{1}{6 z_1} \left[\frac{ 2-e^{-x^*}(2+x^*) }{ r_{cep} }\right]^3 (r-r_{cep})^3 + \nonumber\\
%	 && +o[ (r-r_{cep})^3].
%\end{eqnarray}
%
%
The higher exponent of $M_{3}^*$ with respect to $M_{24}^*$ implies that the critical 2-nodes  are essential for observing the critical end point.
The $k=3$ threshold, in fact, recursively eliminates all the type 2 nodes with exactly two neighbors, triggering a cascade which essentially makes the sub-core become a negligible fraction of the HKC.

%\subsubsection{The critical point}

%\paragraph{Solution of the equations}

The critical point can be calculated by requiring that the maximum ($f_{24}'(Z)=0$) coincides with the second change in convexity ($f_{24}''(Z)=0$).
Hence we get an equation for the critical composition $r_c$:
 \begin{equation}
  \exp\left( 1 + \sqrt{\frac{1-3r_c}{1-r_c}} \right) = 5-7r_c + 4\sqrt{(1-r_c)(1-3r_c)},
 \end{equation}
with  numerical solution  $r_c \approx 0.2962590188$.
Using the appropriate equations we can calculate with arbitrary precision the critical damage $p_c$ and the values of the \kcore~strengths: they are summarized in Table II. %\ref{table:24-critical}.
%~ 
\begin{table}[htb]
\begin{tabular}{ccccccc}
\hline \hline
 $r_c$ & $z_1p_c$ & $z_1M^c_{24}$ & $z_1 C^c_{24}$ & $z_1\bar{M}^c_{2}$ & $z_1\bar{M}^c_{4}$ & $z_1M^c_{3}$ \\
\hline
 0.2963 & 4.1131 & 0.6510 & 0.4079 & 0.4963 & 0.1547 & 0.3569\\
\hline \hline
\end{tabular}
\label{table:24-critical}
\caption{Critical fractions of some relevant quantities in the case $\k=(2,4)$.}
\end{table}

The critical exponent $\beta$ defined by $M_{24}(p) - M_{24}(p_c) \sim (p-p_c)^{\beta}$ can be calculated analytically for each region of the phase diagram:
\begin{equation}
	\beta = \left\{%
	\begin{array}{lll}
		2 & \textrm{critical line} & r>r_{cep}\\
		1/2 & \textrm{hybrid transition} & r<r_c\\
		1/3 & \textrm{hybrid transition} & r=r_c
	\end{array}
	\right.
	\label{eq:2-4-beta-dxt}
\end{equation}
In the region $r_{cep}<r<r_c$, the first order transition separates two percolating phases and therefore it is possible to calculate the exponent $\beta'$ on the left hand side of the first order line (Fig.~\ref{fig:er-2-4}):
\begin{equation}
	\beta' = \left\{%
	\begin{array}{ll}
		1 &  r_{cep}<r<r_c\\
		1/3 & r=r_c
	\end{array}
	\right.
\end{equation}
The exponent $\beta'=1$ away from the critical point is not singular, whereas from the other side of the first order transition it is $\beta=1/2$.
This difference is due to the  presence of the hybrid transition, which is asymmetric.
However, the two exponents coincide at the critical point as generally expected.

It is interesting to note that the exponents at the hybrid transition in (\ref{eq:2-4-beta-dxt}) exactly match the ones found in facilitated spin models reproducing mode-coupling theory singularities \cite{sellitto2012}.
In particular, exponent  $1/2$ corresponds to the $A_2$ singularity associated to a discontinuous liquid-glass transition, whereas the exponent $1/3$ corresponds to the $A_3$ singularity at the endpoint of the discontinuous glass-glass transition of the $F_{13}$ schematic model \cite{sellitto2012,goetze2009}.

The difference $\Delta M_{24}$ of the two coexisting HKC strengths  as $r\to r_c^-$ vanishes according to the following law:
\begin{equation}
	 \Delta M_{24} \sim (r_c - r)^{1/2} .
\end{equation}
Therefore, the associated critical exponent is $\beta_u=1/2$.
The expansion of $\Delta M_{3} $ yields the same critical exponent
\begin{equation}
	 \Delta M_{3} \sim (r_c - r)^{1/2} .
\end{equation}

The phase diagram, then, is characterized by the same topology as in the $k_a=1$, $k_b=3$ case: a first order (hybrid) transition line which ends in a critical point and a critical line which encounters the first order line at a critical end point.
Here the critical point occurs at a higher fraction of low-$k$ nodes, and at a larger fraction of low $k$ nodes in the composition of the HKC: the opposite of the case $\k=(1,3)$.
This is presumably due to the higher fragility of 4-nodes with respect to 3-nodes.

\subsection{The case $k_a=3$, $k_b=8$}

\begin{figure}[htb]
\begin{center}
	%\begin{overpic}[width=0.99\columnwidth,angle=0,grid=5]{figures/3-8-phase-diag-Mr}
	%	\put(60,60){\includegraphics[width=0.3\columnwidth,angle=0]{figures/3-8-phase-diag-pr}}
	%\end{overpic}
	%\showthe\columnwidth % Use this to determine the width of the figure.
	\includegraphics[width=\columnwidth]{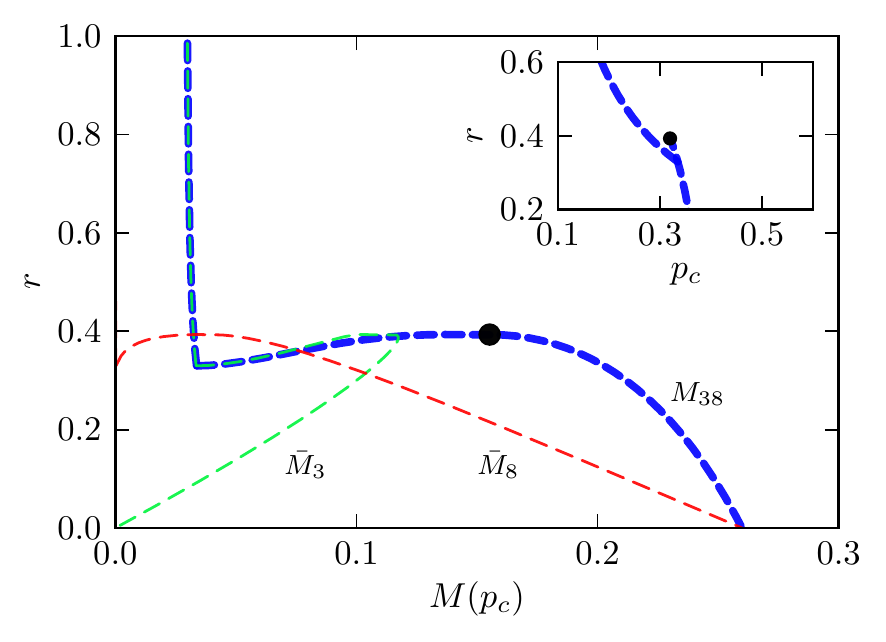}
	\caption{Phase diagram of the case $\k=(3,8)$ for the \er~graph ($z_1=30$). 
	Symbols are as in Fig.~\ref{fig:1-3-ph-diag}.
	Differently from the previous cases, here there are two lines of first order transitions and no classical continuous percolation scenario.}
	\label{fig:er-3-8-phase-diag}
\end{center}
\end{figure}

Binary mixtures involving thresholds higher than 2 cannot present continuous transitions.
However, the mixing of different types of nodes may still result in a critical point.
Fig.~\ref{fig:er-3-8-phase-diag} shows the phase diagram of the case $\k=(3,8)$ as an example of this behavior.
The high difference in resilience between 3-nodes and 8-nodes results in an intermediate region of the phase diagram where an 8-rich phase collapses into a 3-rich phase.
The two lines of first order transitions do not match smoothly, and a region with a stable 3-rich percolating phase can be easily seen.
It can be shown that the critical point is in the same  universality class as the one in the case $\k=(2,4)$, with the same critical exponents along the first order line.

%\begin{figure}[htb]
%\begin{minipage}[]{0.49\columnwidth}
%	\includegraphics[width=1.1\columnwidth,angle=0]{figures/3-8-phase-diag-Mr}
%\end{minipage}
%\begin{minipage}[]{0.49\columnwidth}
%	\includegraphics[width=1.1\columnwidth,angle=0]{figures/3-8-phase-diag-pr}
%\end{minipage}
%\caption{Phase diagram of the case $\k=(3,8)$ for \er~graphs ($z_1=30$).}
%\label{fig:er-3-8-phase-diag}
%\end{figure}

\subsection{Summary}
\begin{figure}[htb]
	\includegraphics[width=\columnwidth]{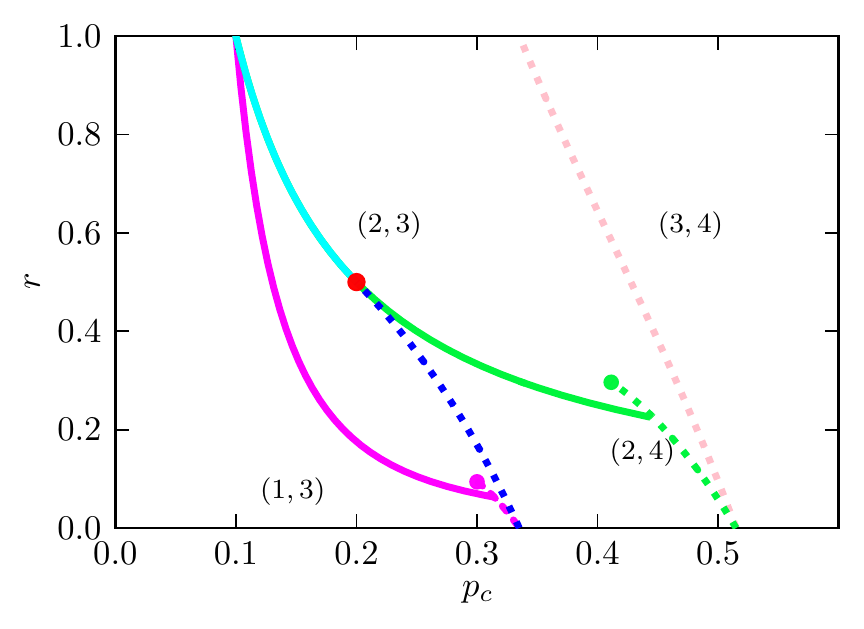}
	\caption{
		Comparison of the phase diagrams of the cases $\k=(1,3)$, $\k=(2,3)$, $\k=(2,4)$ and $\k=(3,4)$ for \er~graphs with $z_1=10$.
		As in the previous phase diagrams, continuous lines represent continuous phase transitions, whereas dashed lines represent first order phase transitions.
		The red dot represents a TCP; the other dots are critical points.}
	\label{fig:comparison-phase-diag-pr}
\end{figure}

Fig.~\ref{fig:comparison-phase-diag-pr} summarizes the phase diagram of a few relevant binary mixtures.
The general behavior of mixtures of type $\k=(1,k)$ or $\k=(2,k)$ is characterized by a critical line and a line of first order transitions which ends in a critical point.
The case $\k=(2,3)$, however, is quite peculiar as the two lines match at a tricritical point.
This occurrence is not obvious and will be investigated in the next section.
Both the tricritical and the critical point observed in most mixtures belong to the same universality class of facilitated spin models reproducing mode coupling theory singularities of type $F_{12}$ and $F_{13}$, respectively \cite{sellitto2012,arenzon2012}.

Fig.~\ref{fig:comparison-phase-diag-pr} also reports the case $\k=(3,4)$ as an example where there cannot be continuous transitions, as both values of $k$ are larger than 2.
The mixture $\k=(3,4)$ is only characterized by a line of first order transitions.
As we have seen in the case $k=(3,8)$, though, a critical point can still arise when the two values of $k$ are quite far from each other so that the high-$k$ phase can collapse without undermining the stability of a low-$k$ phase (Fig.~\ref{fig:er-3-8-phase-diag}).

\section{A criterion for a tricritical point}

We have already noted that the occurrence of a TCP is quite peculiar in this type of models, but quite resilient to different network topologies \cite{cellai2011}.
In this section we give some insight into the origin of a TCP and show a criterion to establish the occurrence of it in a given binary mixture of vertices. 
As a general rule, given a binary mixture $\k=(k_a,k_b)$ where $k_a$- and $k_b$-cores are characterized by a second and a first order transition, respectively, the TCP occurs whenever the critical point associated with to the discontinuous transition of the $k_b$-rich mixture coincides with a vanishing HKC strength.
The critical percolation line, by definition, is characterized by a vanishing HKC and must match the first order line to give origin to a TCP.

In the following we assume that the degree distribution $P(q)$ satisfies
\begin{equation}
	0 <\sum_{q=3}^{+\infty}  q(q-1)(q-2)P(q)<+\infty.
\end{equation}
This condition is quite general, as it has already been shown that scale free networks with $P(q)\sim q^{-\gamma}$ (for large $q$) and $1<\gamma\le 3$ are not characterized by first order transitions and therefore their phase diagrams cannot involve neither critical or tricritical points \cite{dorogovtsev2006}.
Therefore, the only interesting case where the following analysis does not apply involves scale free networks with $3<\gamma\le 4$.

Due to the different properties of binary mixtures, we differentiate  two cases according to the presence of finite HKC clusters or not.

\subsection{The case $k_a=2$, $k_b=k\ge 3$}

In the case $\k=(2,k)$, equation (\ref{eq:Z-equation}) can be re-written as $pf_{2k}(Z) = 1$, where   the relevant function is
\begin{IEEEeqnarray}{rCl}
	f_{2k}(Z) &=& r\sum_{q=2}^{+\infty} \frac{qP(q)}{\langle q \rangle} \sum_{l=1}^{q-1}  {q-1 \choose l} Z^{l-1}(1-Z)^{q-1-l} +\nonumber\\
		&& +(1-r)\sum_{q=k}^{+\infty} \frac{qP(q)}{\langle q \rangle} \sum_{l=k-1}^{q-1}  {q-1 \choose l} Z^{l-1}(1-Z)^{q-1-l}.\nonumber\\
\end{IEEEeqnarray}
In this case $X=Z$, so we only need to study the behaviour of the equation at $Z\to 0^+$.
The expansion reads
\begin{IEEEeqnarray}{rCl}
	f_{2k}(Z) &=& r\sum_{q=2}^{+\infty} \frac{q(q-1)P(q)}{\langle q \rangle} +\nonumber\\
	&&	-\frac{1}{2}r \sum_{q=2}^{+\infty} \frac{q(q-1)(q-2)P(q)}{\langle q \rangle}Z +O(Z^2) +	\nonumber\\
		&& +(1-r)\sum_{q=k}^{+\infty} \frac{qP(q)}{\langle q \rangle} {q-1 \choose k-1} Z^{k-2} + O(Z^{k-1}).\nonumber\\
		\label{eq:f2k-expansion}
\end{IEEEeqnarray}
A necessary condition for  a critical point which continuously approaches the line $Z=0$ for $r\to r_c$ is that $f_{2k}'(0) = 0$ at $r=r_c$.
From equation (\ref{eq:f2k-expansion}), we have two possibilities. 
If $k\ge 4$,
\begin{equation}
	f_{2k}'(0) = 	-\frac{1}{2}r \sum_{q=2}^{+\infty} \frac{q(q-1)(q-2)P(q)}{\langle q \rangle},
\end{equation}
which is always negative and vanishes only for $r=0$.
If $k=3$, instead, the linear term proportional to $r$ mixes with the linear term from the type 3 nodes and we get
\begin{equation}
	f_{23}'(0) = \left(\frac{1}{2}-r\right) \sum_{q=3}^{+\infty} \frac{q(q-1)(q-2)P(q)}{\langle q \rangle},
	\label{eq:fprime23}
\end{equation}
which vanishes at $r=1/2$.

From equation~(\ref{eq:f2k-expansion}) it transpires that $Zf_{2k}(Z)$ can be written as  an expansion in terms of core shells for each node type.
The condition of the critical point $f_{2k}'(Z)=0$ implies that the dominant term for $Z\to 0^+$ is the linear one.
The term $r\sum_{q} \frac{q(q-1)(q-2)P(q)}{2 \langle q \rangle}Z^2$ represents the probability that, following an edge leading to a node of type 2, there are two outgoing edges connected to the HKC (\ie~there must be exactly one extra neighbor beyond  the minimum possible).
The term $(1-r)\sum_{q} \frac{qP(q)}{\langle q \rangle} {q-1 \choose k-1} Z^{k-1}$ represents the probability that, following an edge leading to a node of type $k$, there are $k-1$ outgoing edges connected to the HKC.
In other words, this is the probability that the $k$-node found at the end of the edge is connected to the $k$-corona.
The condition for a TCP, therefore, is that the $3$-shell term of the type 2 nodes has the same order of magnitude as the $k$-corona of the $k$ type nodes.
This can only occur when $k=3$, where we have a TCP at $r=1/2$ (see also Eq.~\ref{eq:fprime23}).

\subsection{The case $k_a=1$, $k_b=k\ge 3$}

As noticed in Section 3, finite \kcore s exist when $k=1$ \cite{cellai2011}.
Therefore, we have $X<Z$, and the functions associated to the two relevant equations (\ref{eq:Z-equation}) and (\ref{eq:X-equation}) are:
\begin{equation}
	f_{1k}(Z) = \frac{r}{Z} + (1-r)\sum_{q=k}^{+\infty} \frac{qP(q)}{\langle q \rangle} \sum_{l=k-1}^{q-1}  {q-1 \choose l} Z^{l-1}(1-Z)^{q-1-l},
	\label{eq:f1k}
\end{equation}
\begin{IEEEeqnarray}{rCl}
	h_{1k}(X,Z) &=& r \sum_{q=0}^{+\infty} \frac{q P(q)}{\langle q \rangle} \sum_{m=1}^{q-1} {q-1 \choose m} X^{m-1} (1-X)^{q-1-m}+	\nonumber\\
		&& +(1-r)\sum_{q=k}^{+\infty} \frac{qP(q)}{\langle q \rangle} \sum_{l=k-1}^{q-1} {q-1 \choose l} (1-Z)^{q-1-l} \times \nonumber\\
		&& \times \sum_{m=1}^{l} {l \choose m} X^{m-1} (Z-X)^{l-m}.
	\label{eq:h1k}
\end{IEEEeqnarray}
As in the case $\k=(2,k)$, it is meaningful to expand $h(X,Z)$ for  $X\to 0^+$:
\begin{equation}
	h_{1k}(X,Z) = a_0(Z) + a_1(Z)X + O(X^2),
	\label{eq:hXZ-expansion-1k}
\end{equation}
where
\begin{IEEEeqnarray}{rCl}
	a_0(Z) &=& r\sum_{q=2}^{+\infty} \frac{q(q-1)P(q)}{\langle q \rangle} +(1-r) \sum_{q=k}^{+\infty}\frac{qP(q)}{\langle q \rangle} \times\nonumber\\
	&& \times  \sum_{l=k-1}^{q-1}  {q-1 \choose l}l Z^{l-1}(1-Z)^{q-1-l};
\end{IEEEeqnarray}
\begin{IEEEeqnarray}{rCl}
	a_1(Z) &=& -\frac{1}{2}r \sum_{q=2}^{+\infty} \frac{q(q-1)(q-2)P(q)}{\langle q \rangle} +\nonumber\\
	&& -\frac{1}{2}(1-r) \sum_{q=k}^{+\infty} \frac{qP(q)}{\langle q \rangle}\sum_{l=k-1}^{q-1} {q-1 \choose l}\times\nonumber\\
	&&	\times  l (l-1) Z^{l-2}(1-Z)^{q-1-l}.
\end{IEEEeqnarray}
It is  important to remark that $a_1(Z)<0$ for every $Z$, \ie~$h(0,Z)$ is always a local maximum of $h$ as a function of $X$.

%\subsubsection{A necessary condition for a tricritical point}

If a TCP exists, there must exist a value of $r$ for which a critical point (the endpoint of a line of first order transitions) occurs exactly at $X=0$, as this corresponds to the critical line.
As $\lim_{Z\to 0}f_{1k}(Z) = +\infty$ for every $r>0$ (Eq.~\ref{eq:f1k}), the equation $pf_{1k}(Z)=1$  never has a solution $Z=0$ and $f_{1k}$ is monotonic in the neighborhood of $Z=0$.
Therefore, a line of first order transitions corresponds to a local maximum of $f_{1k}$, where equation $pf_{1k}(Z)=1$ has two solutions.
The discontinuity in the solution $Z(p)$ provokes a discontinuity in the solution $X(p)$ of the second equation.
This in turn causes a discontinuity in the strength $\S$ of the giant HKC.
Therefore, if a critical point occurs at $Z=Z_c$, it must be $f_{1k}'(Z_c)=f_{1k}''(Z_c)=0$.
(The local maximum of $f$ cannot occur at $Z=0$, as we have seen.)
In order to have a TCP, the above defined critical point must occur at $X_c=0$.
This implies that the fraction $p_c$ of undamaged nodes at criticality must satisfy the inequality
\begin{equation}
	h_{1k}(0,Z_c)\le \frac{1}{p_c}.
\end{equation}
Using the first equation, this is equivalent to
\begin{equation}
	h_{1k}(0,Z_c)\le f(Z_c),
	\label{eq:TCP-condition}
\end{equation}
where $f_{1k}'(Z_c)=f_{1k}''(Z_c)=0$ and, in case of multiple solutions, the value of $Z_c$ where $f_{1k}(Z_c)$ is highest must be considered.

The condition (\ref{eq:TCP-condition}) is only a necessary condition for a TCP, because in general it may be possible that $h$ has a maximum higher than $h_{1k}(0,Z)$.
However, we now show that this condition is violated for all values of $k$.
Let us consider the two conditions:
\begin{equation}
	\begin{cases}
		h_{1k}(0,Z)\le f_{1k}(Z) \\
		f_{1k}'(Z)=0
	\end{cases}.
\end{equation}
Substituting the second equation into the first one yields
\begin{IEEEeqnarray}{rCl}
	\lefteqn{r\sum_{q=2}^{+\infty} \frac{q(q-1)P(q)}{\langle q \rangle} Z +} \nonumber\\
	&&+  (1-r) \sum_{q=k}^{+\infty} \frac{qP(q)}{\langle q \rangle} \sum_{l=k}^{q-1}  {q-1 \choose l}l Z^{l}(1-Z)^{q-1-l} \le 0
	\label{eq:TCP-condition-1k}\nonumber\\
\end{IEEEeqnarray}
which is never true, because both sums on the left hand side are strictly positive for any $Z>0$.

It is interesting to note that the expansion  (\ref{eq:hXZ-expansion-1k}) has terms which are reminiscent of a $k$-shell decomposition (as in Eq.~(\ref{eq:f2k-expansion})).
The violation of the TCP condition (\ref{eq:TCP-condition-1k}) appears to be related to a mis-match between 2-shells of 1-nodes and the leading term in $Z$ of $k$-nodes.

Both in the cases $\k=(2,k)$ and $\k=(1,k)$, then, we can say that a necessary condition for the occurrence of a TCP is the vicinity of the two values of $k$.
In the case $\k=(1,k)$, a TCP never occurs, whereas in the case $\k=(2,k)$ it is only present for $k=3$, where the 3-shell of type 2 vertices has the same order of magnitude as the corona of type 3 vertices.

\section{Conclusions}

%What does  the TCP criterion say about FSPs?

In this paper, we have explored and classified all the critical phenomena which characterize binary mixtures in heterogeneous \kcore~percolation.
The most interesting critical phenomena involve mixtures which separately give origin to phase transitions of different order.
In this set of cases, we observe two main phase diagram topologies: a critical line intercepted by a first order line ending in a critical point, or a critical line matching the first order line onto a tricritical point (TCP).
A careful analysis of equations (\ref{eq:Z-equation}) and (\ref{eq:X-equation}) for different types of binary mixtures shows that the occurrence of a TCP scenario is strictly related to the relative order of magnitude of $(k_a+1)$-shells and the $k_b$-corona (when $k_a<k_b$) in approaching the hybrid transition.
Broadly speaking, this condition quantifies the closeness of the parameters $k_a$ and $k_b$ in order for a TCP to occur.
This criterion may be potentially important in network engineering (for example in managing large scale infrastructures), as the presence of a TCP implies a smooth change in the way the network collapses, and the possibility to control the order of the transition.

This behavior is also reminiscent of the scenario observed in binary mixtures of hard spheres, where a discontinuous glass-glass transition, ending into a critical point, weakens up to disappearing when the sizes of the two particle types become similar \cite{voigtmann2011}.
The analogy is even more striking, as recently proposed facilitated spin models, which belong to the same universality class of this HKC model \cite{sellitto2010}, show with increasing evidence the correct reproduction of high-order singularities of the mode-coupling theory \cite{arenzon2012,cellai2013} and a strong link with kinetically constrained models \cite{foini2012}.
In other words, high and low $k$s in \kcore~percolation may be associated to large and small particles, respectively, as a large particle requires on average more neighboring particles to become caged and vice-versa for small particles.

In summary, heterogeneous \kcore~percolation provides an exactly solvable model characterized by a wealth of interesting critical phenomena.
The model has a great potential in investigating the fundamental processes occurring in the proximity of critical points and therefore in giving insights into quite different interesting applications.

%What is the difference between MCT singularities?

%What is the meaning of corona clusters in FSMs?

We acknowledge useful conversations with Gareth Baxter and Mauro Sellitto.
This work has been partially funded by Science Foundation Ireland, grants 11/PI/1026 and 06/MI/005, and the FET-Proactive project PLEXMATH (FP7-ICT-2011-8; grant number 317614) funded by the European Commission.

%\bibliography{library}


\end{document}